\documentclass[conference]{IEEEtran}
\IEEEoverridecommandlockouts
\usepackage{cite}
\usepackage{amsmath,amssymb,amsfonts}
\usepackage{algorithmic}
\usepackage{graphicx}
\usepackage{textcomp}
\usepackage{xcolor}

\usepackage{hyperref}

\def\BibTeX{{\rm B\kern-.05em{\sc i\kern-.025em b}\kern-.08em
    T\kern-.1667em\lower.7ex\hbox{E}\kern-.125emX}}
\begin{document}

\title{A Deep Learning-Based Particle-in-Cell Method for Plasma Simulations
}

\author{\IEEEauthorblockN{Xavier Aguilar, Stefano Markidis}
\IEEEauthorblockA{
\textit{Electrical Engineering and Computer Science School}\\
\textit{KTH Royal Institute of Technology}\\
Stockholm, Sweden \\
}
}

\maketitle

\begin{abstract}
We design and develop a new Particle-in-Cell (PIC) method for plasma simulations using Deep-Learning (DL) to calculate the electric field from the electron phase space. We train a Multilayer Perceptron (MLP) and a Convolutional Neural Network (CNN) to solve the two-stream instability test. We verify that the DL-based MLP PIC method produces the correct results using the two-stream instability: the DL-based PIC provides the expected growth rate of the two-stream instability. The DL-based PIC does not conserve the total energy and momentum. However, the DL-based PIC method is stable against the cold-beam instability, affecting traditional PIC methods. This work shows that integrating DL technologies into traditional computational methods is a viable approach for developing next-generation PIC algorithms.
\end{abstract}


\section{Introduction}
Particle-in-Cell (PIC) methods are among the most powerful and used computational methods for the simulation of plasmas with application to fusion reactors, laser-plasma devices, accelerators, space physics, and astrophysics. Massively parallel PIC codes such as VPIC~\cite{bowers2009advances}, iPIC3D~\cite{markidis2010multi}, and Warp-X~\cite{vay2018warp} daily run on the largest supercomputers in the world. The PIC simulations analysis vastly contributed to advancing our understanding of plasma dynamics in complex phenomena and systems.

The basic PIC methodology was developed during the early Eighties by Harlow, Dawson, Buneman, Birdsall, Langdon, and many others. In essence, the PIC method follows the trajectories of billion electrons and protons in a self-consistent electromagnetic field, determined by solving Maxwell's equations. In recent decades, several algorithmic advancements, such as semi-implicit and fully implicit methods \cite{markidis2011energy,markidis2018polypic}, coupled fluid-kinetic approaches \cite{markidis2014fluid,markidis2018polypic}, have largely advanced the state-of-the art and the applicability of PIC methods to study very large systems, such as large planetary magnetospheres~\cite{chen2017global}, over long simulations. 

Recently, Machine Learning and Deep Learning (DL) methods have emerged as valuable tools for data analysis and to replace or complement more traditional computational approaches. An example of such efforts is the development of DL-based preconditioners and linear solvers~\cite{lunaaccelerating,ichimura2020fast}, heterogeneous linear solvers for Partial Differential Equations~\cite{markidis2021physics}, use of DL-based methods in Computational Fluid Dynamics (CFD) solvers~\cite{guo2016convolutional}, or weather forecasting (tropical cyclone estimation)~\cite{pradhan2017tropical}. Furthermore, DL has gained prominence within the scientific community and within the core High-Performance Computing (HPC) community. For example, the HPC community investigated DL frameworks on HPC systems~\cite{shams2017evaluation,chien2018characterizing}, the adaptation of DL methods to run natively on HPC clusters~\cite{yoginath2019towards, chien2019tensorflow}, and the DL usage to predict system health and time to failure in HPC~\cite{das2018desh}. 

The goal of this study is two-fold. The first goal is to design and develop a new PIC method employing DL technologies in the PIC computational cycle. The second goal is to determine the advantages of DL-based PIC methods compared to traditional PIC methods in terms of stability and accuracy. 

The main contributions of this work are the following:
\begin{enumerate}
\item We design and develop a methodology to embed a DL electric field solver into the PIC method. The DL electric field solver is trained using particle phase space information and associated electric field.
\item We compare the accuracy and the performance of the DL-based PIC method with the traditional PIC method, showing that the DL-based PIC method reproduces the correct results for the two-stream instability with acceptable total energy and momentum variation.
\end{enumerate}
Overall, we show that integrating DL technologies in PIC methods is a promising approach to extending and going beyond existing PIC methods.

The paper is organized as follows. We present first the basic formulation of a traditional PIC method in Section~\ref{background}. Section~\ref{method} describes the algorithm and design of the DL-based PIC method for solving the two-stream instability problem. We detail the experimental set-up, such as the DL network architectures and data sets, in Section~\ref{setup}. Section~\ref{result} presents the results obtained with the DL-based PIC method and compares them with the results of a traditional PIC method. We briefly discuss previous work in Section~\ref{related} and summarize the paper and outline future work in Section~\ref{conclusion}.

\section{Background}
\label{background}
\begin{figure}[tb]
\centering
\centerline{\includegraphics[width=\columnwidth]{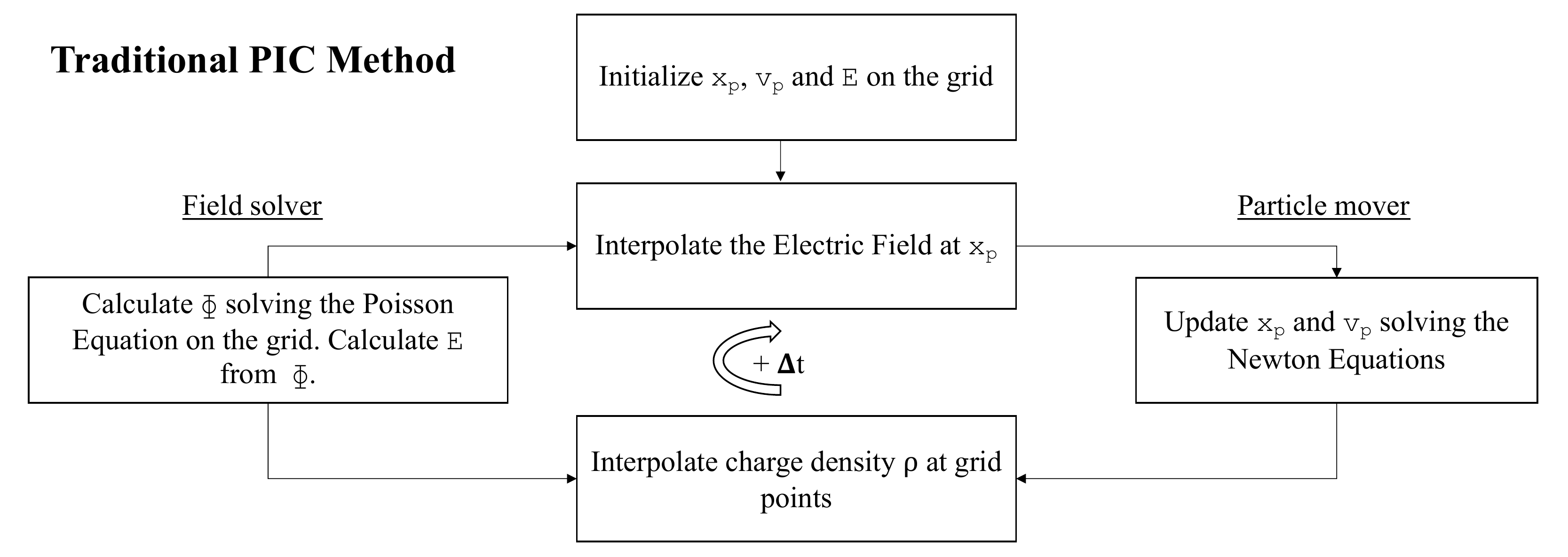}}
\caption{Traditional PIC algorithm.}
\label{img_trad}
\end{figure}

The goal of this work is the development of a new PIC method employing DL technologies. In this work, we call the explicit-in-time PIC method \emph{traditional PIC method}~\cite{birdsall2018plasma}. The traditional PIC algorithm is depicted in Fig.~\ref{img_trad}.

The basic PIC algorithm consists of an initialization phase and a repeated computational cycle hundreds or thousands of times. 

In the first initialization phase, we set the values of the particle (electron and proton) positions and velocities and the electric and magnetic field values on cells. For instance, we can initialize particle positions uniformly in space and particle velocities with Gaussian distribution(with mean velocity $v_0$ and thermal spread $v_{th}$). The initial field is initialized consistently with values of charge and current densities, satisfying Maxwell's equations.

After the initialization, a computational cycle is repeated for several iterations. The computational step is divided into four phases. 

First, the electric field and magnetic fields (defined at the grid cells) are calculated at the particle positions. This task is carried out by extrapolating the field's values at interpolation using constant (Nearest Grid Point, NGP) or linear (Cloud-in-Cell, CIC) or higher-order interpolation functions~\cite{birdsall2018plasma}.  

The second step is the particle mover or pusher. In this stage, particle positions ($x_p$) and velocities ($v_p$) are updated by solving the Newton equation. Using a simple leap-frog scheme in one dimension and the electrostatic regime without a magnetic field~\cite{birdsall2018plasma}:
\begin{equation}
\label{xp}
x_p^{n+1} = x_p^n + v_p^{n+1/2} \Delta t 
\end{equation}
\begin{equation}
\label{vp}
v_p^{n+1/2} = v_p^{n-1/2} + q / m E_p^{n} \Delta t 
\end{equation}
where $n$ is the time level, $\Delta t$ is the simulation time step, $q/m$ is the charge over mass ratio and $E_p$ is the electric field acting on the particle.

The third step is the calculation of charge density and current densities on the grid cells. In the basic NGP scheme, the charge density for a cell is obtained by measuring how many particles are located in the cell, multiplying this number by the particle charge and dividing it by the cell volume. This step is also called \emph{interpolation} and higher-order interpolation functions can be used.

The fourth step is the so-called \emph{field-solver} stage that solves Maxwell's equations on a grid. In the case of electrostatic limit, Maxwell's equations reduce to only solve the Poisson equation. We calculate the electrostatic potential ($\Phi$) solving the Poisson equation given the charge density ($\rho$) from the previous interpolation step:
\begin{equation}
\nabla^2 \Phi =  - \rho / \epsilon_0 ,
\end{equation}
where $\epsilon_0$ is vacuum permittivity. This PDE is typically solved by using a finite difference numerical scheme that requires the solution of a linear system. Once the electrostatic field, $\Phi$ is known, the electric field $E$ is calculated as:
\begin{equation}
E = - \nabla \Phi 
\end{equation}

The gradient operator is discretized on the PIC grid and the E is solved by finite difference from the $\Phi$ values.

Several PIC methods have been developed using different discretization schemes of the PIC governing equations. PIC methods are characterized by different numerical properties in terms of stability and conserved quantities, such as total energy and momentum. While these quantities are conserved in real physical systems, they are typically not in modeled systems as numerics introduces numerical artifacts. For instance, traditional explicit PIC methods are conditionally stable and momentum conserving~\cite{birdsall2018plasma}.  Fully-implicit PIC schemes are numerically stable and can conserve the total energy of the system~\cite{markidis2011energy}.

\section{Methodology}
\label{method}
For simplicity and demonstration purposes, we focus on a well-known plasma benchmark test, the two-stream instability. In line with previous studies, we use a one-dimensional geometry. We study the instability in the electrostatic limit in the absence of a magnetic field. We also assume that the protons are motionless in the background to neutralize the plasma. The two-stream instability occurs over a faster time than the massive proton dynamics (protons mass is 1836 times larger than electrons, and therefore, their inertia is much higher than electrons). We also fix the box size ($L$) equal to $2 \pi / 3.06 $. This size is chosen to accommodate the most stable mode for two beams drifting at average velocity $v_0 = \pm 0.2$. We also fix the number of cells in the PIC simulation to 64, the number of electrons to 1,000 per cell and the simulation time step to 0.2. All the PIC quantities in this work are dimensionless with the electron $q/m$ equal to one.

\begin{figure}[tb]
\centering
\centerline{\includegraphics[width=\columnwidth]{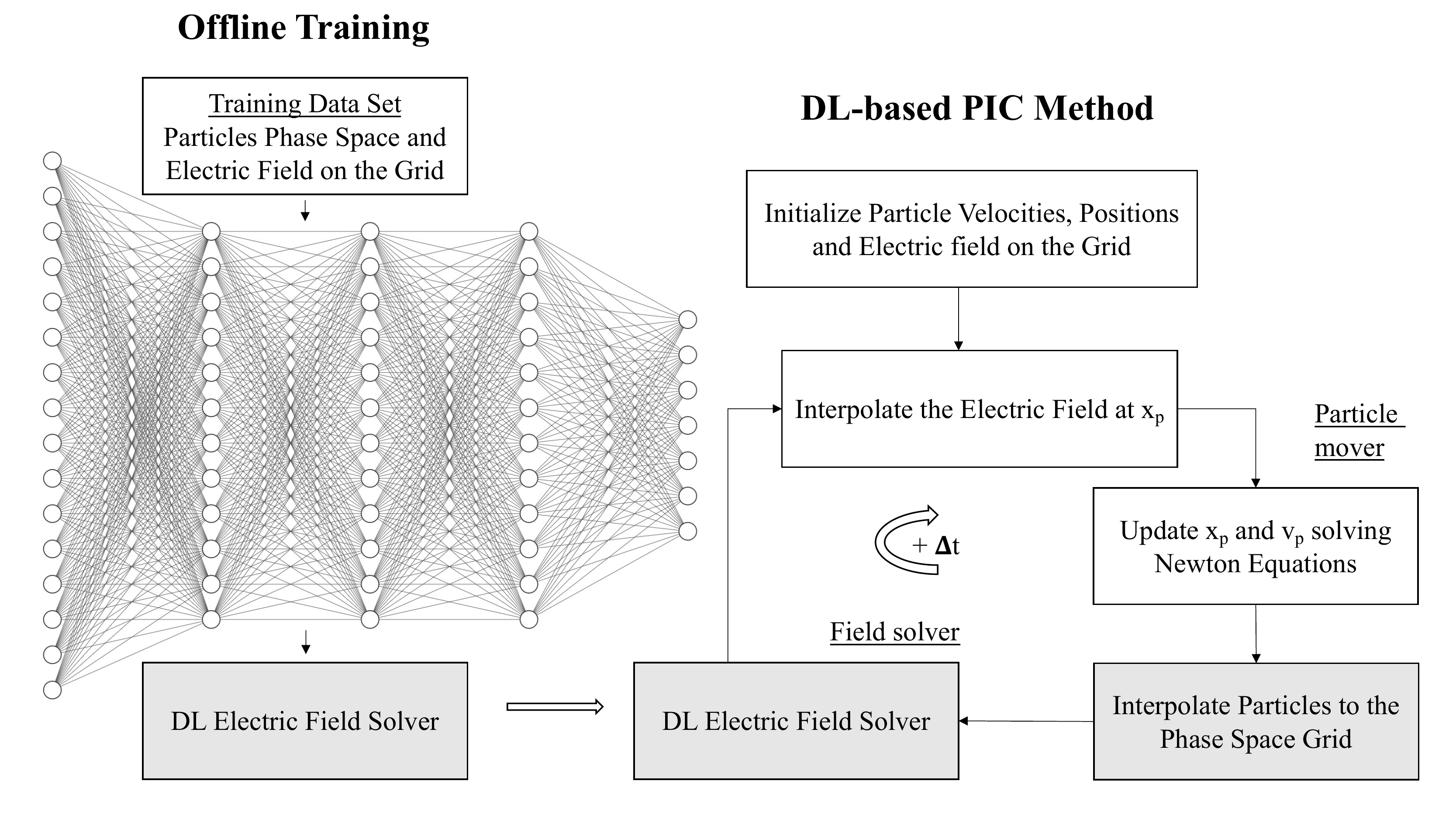}}
\caption{The DL-based PIC method introduces a binning of the phase space and DL electric solver.}
\label{img_DLpicAlgo}
\end{figure}

A diagram of the DL-based PIC method is presented in Fig.~\ref{img_DLpicAlgo}. The DL PIC method still retains the interpolation step to calculate the electric field at the particle position and particle mover (Eqs.~\ref{xp} and~\ref{xp}). The DL-based PIC method replaces the interpolation step for the charge and densities calculations and electric solver of the traditional PIC method with two new steps (in grey color in Fig.~\ref{img_DLpicAlgo}): an interpolation of particle velocity and position into a phase space grid and a DL electric field solver that is the result of a DL neural network training. 

The phase space comprises all particle velocities and positions. In a 1D simulation, the phase space is represented by depicting each particle as a point on a scatter plot with the position as $x$-coordinate and velocity as $y$-coordinate. We form a phase space grid by discretizing phase space with a two-dimensional grid and counting how many particles belong to a cell of the phase space grid.

\begin{figure}[tb]
\centering
\centerline{\includegraphics[width=\columnwidth]{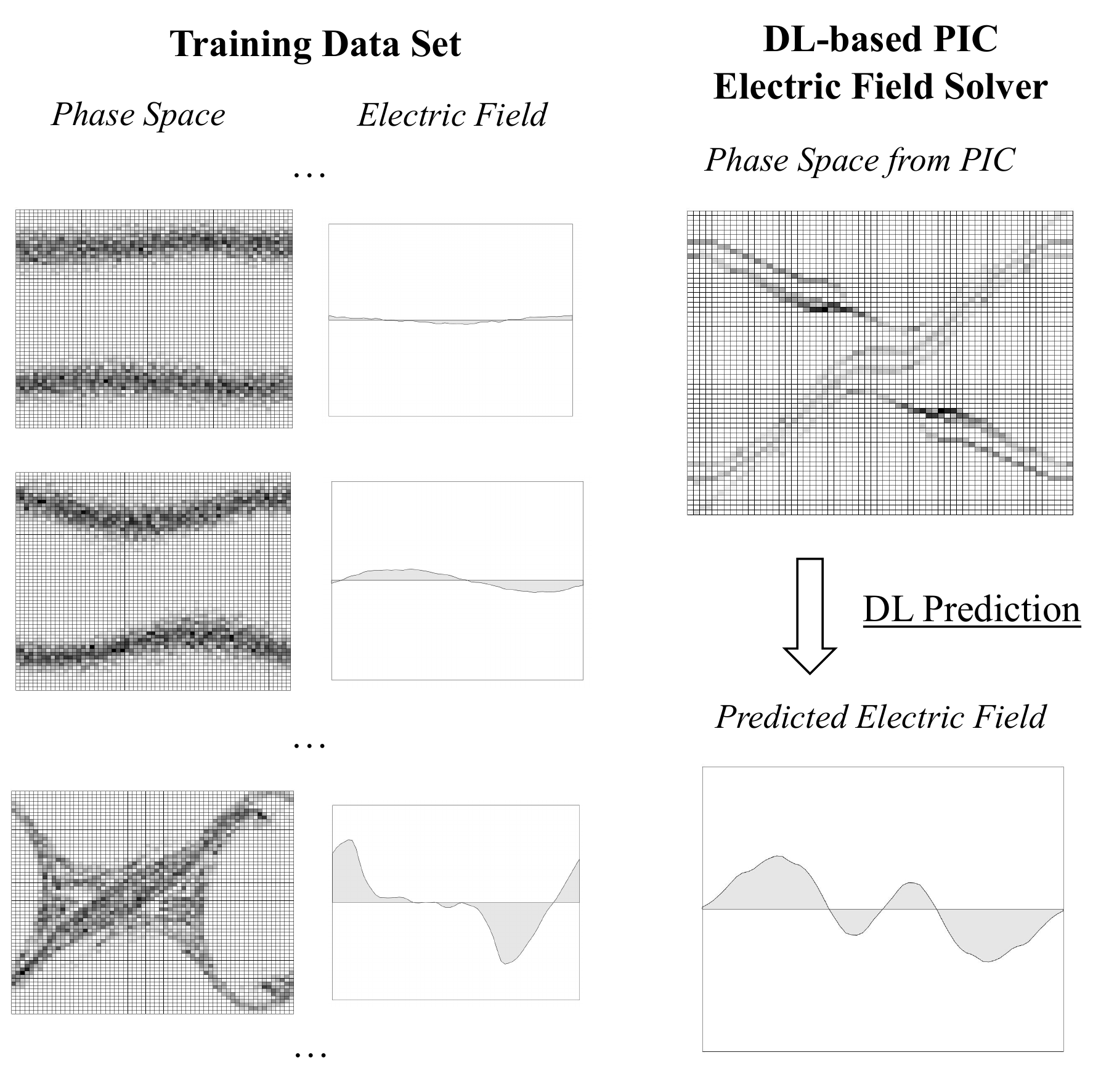}}
\caption{The training set consists of electron phase space and associated electric field. The DL-based PIC method predicts the electric field taking the phase space information as input at each PIC cycle.}
\label{img_DLpic}
\end{figure}

After the phase space binning, we use the DL electric field solver to calculate (or predict) the electric field given a phase space grid (as shown in the right panel of Fig.~\ref{img_DLpic}). The DL electric field solver is obtained by training a DL neural network.

To train the DL electric solver, we produce a training data set, formed by phase space grid and associated electric field. Examples of phase space grid and associated electric field are shown in the left panel of Fig.~\ref{img_DLpic}. The training set is produced by running highly accurate traditional PIC simulations. For instance,  we can use relatively small steps and high-order interpolation functions.

\section{Experimental Set-up}
\label{setup}
In this section, we describe the setup we use for validating the DL-based PIC method.

\subsection{Neural Network Architecture}
The major design choice for the DL electric solver is the selection of the neural network architecture. Several choices on the architecture are possible: for instance, it can be a generic Multilayer Perceptron (MLP), a Convolutional Neural Network (CNN) or a Residual Network (ResNet), among others. In this paper, we explore the use of DL within a PIC code using an MLP and a CNN. 

First, we investigate the performance of an MLP network with three hidden layers. Each hidden layer is fully connected and contains 1,024 neurons with a \textit{Relu} activation function. The output layer consists of 64 neurons with a \textit{Linear} activation function, because we want to learn a multi-variate regression function of the electric field on 64 cells. 

Second, we test our DL-based PIC method using a more complex neural network architecture. In this case, we implement a network with two blocks of convolutional layers followed by three fully connected layers. Each convolutional layer block was composed of two convolutional layers followed by a MaxPooling layer. The three fully connected layers follow the same configuration than in the simple model, 1024 neurons each with a \textit{Relu} activation function.
Finally, the output layer is the same as in the MLP, 64 neurons with a \textit{linear} activation.

We used the Adam optimizer with a batch size of 64 samples and a learning rate of 0.0001. 

\subsubsection{Training, Test \& Validation Data Sets}
Our data set consists of 40,000 images and 40,000 text files generated from several traditional PIC simulations using combinations of the initial beam velocities ($\pm v_0$) and the thermal speed ($v_{th}$). More concretely, we collected data for 
20 combinations of these two parameters, being $v_0 = [ \pm 0.05, \pm 0.15, \pm 0.18, \pm 0.1, \pm 0.3 ]$ and 
$v_{th} = [0.0, 0.01, 0.001, 0.005]$. For each single combination we collected data from 10 experiments 
(traditional PIC simulations) as a way of data augmentation, since each experiment have similar properties but
slightly different numerical values. Furthermore, we run 200 time steps in each traditional PIC simulation (after 200 time steps the two-stream instability is fully developed in all the PIC simulations), and for each time-step we generated a discretized phase-space 2D-histogram and a text file containing the values 
of the electric field. Fig.~\ref{img_DLpic} shows some examples of such phase-space histograms together with 
their corresponding representation of the electric field. We inspected all the data sets to ensure that no numerical 
instability or artifacts were present.

The total size of our complete data set is 5.2GB, being 
267MB per test case (combination), and 27MB per experiment. The data set was shuffled and then divided into 38,000 
images for training, 1,000 images for validation, and 1,000 images for testing. 

Finally, we also generated an additional test set (called Test Set II) with 1,000 samples from simulations using parameters not included in the initial data set. The main reason is the nature of our experiments and the data augmentation performed. Since our initial data set contained samples from 10 experiments for each combination, many samples in the initial test set were probably slightly similar to other instances in the training set. Therefore, we measure the performance with a data set containing samples that the networks had not seen before. 

The input data, i.e., the phase space grid, were normalized before being fed into the network. All their values were transformed from their original range to $[0,1]$ using the formula below:

\begin{equation}
   y = \dfrac{x - min}{max - min},
\end{equation}
where $min$ and $max$ are the minimum and maximum values in the data set.

We used 150 and 100 epochs for the training of the MLP and CNN network, respectively.

\subsubsection{Hardware and Software Environment}
We run our experiments in a node with two 12-core Intel E5-2690V3 Haswell processors. The computational node has 512 GB of RAM and one Nvidia Tesla K80 GPU card, also used in the training of the networks. The neural networks are implemented using TensorFlow \cite{tensorflow2015-whitepaper} and Keras \cite{chollet2015keras}. On such a system, the training of the MLP network and CNN take approximately 18 minutes and 2 hours, respectively. 

\section{Results}
\label{result}
In this section, we show the results of the DL electric solver first and then DL-based PIC method employing the DL electric field solver as part of the main computational cycle.

As main metric for the performance of the DL-based electric field solver, we use the Mean Absolute Error (MAE) to evaluate the prediction error: 
\begin{equation}
\dfrac{\sum_{1=1}^{n_s}|E_{pred,i} - E_{i}|}{n_s} ,
\end{equation}
where $i$ is the sample in the data set and $n_s$ is the number of total samples.

Table~\ref{table_MAE} shows the MAE for each architecture with our two test sets, that is, \emph{set I} that contains samples 
using same parameters as in the training set, and \emph{set II}, which includes samples from simulations using parameters not 
used in the training set. As a reference, the maximum electric field value obtained in the simulations is approximately 0.1. The results show that the MLP architecture exhibits better performance than the CNN for both test sets (lower MAE corresponds to better performance).
The table also shows that for the CNN, as expected, the performance is worse with the set II, because it contains 
samples from simulations using $\pm v_0$ and $v_{th}$ values not seen before by the network. However, that is not the case
with the MLP, and therefore, further investigation to explain this phenomenon is needed. 

\begin{table}[!t]
\renewcommand{\arraystretch}{1.3}
\caption{MAE and maximum error with each network}
\label{table_MAE}
\centering
\begin{tabular}{|c||c||c|c|}
\hline
\bfseries Metric & \bfseries Test Set & \bfseries MLP & \bfseries CNN\\
\hline\hline
Mean Absolute Error & I & 0.0019 & 0.0020\\
Max Error & I & 0.06899 & 0.0463\\
\hline
Mean Absolute Error & II & 0.0015 & 0.0032\\
Max Error & II & 0.0286 & 0.073\\
\hline
\end{tabular}
\end{table}
\begin{figure}[tb]
\centering
\centerline{\includegraphics[width=\columnwidth]{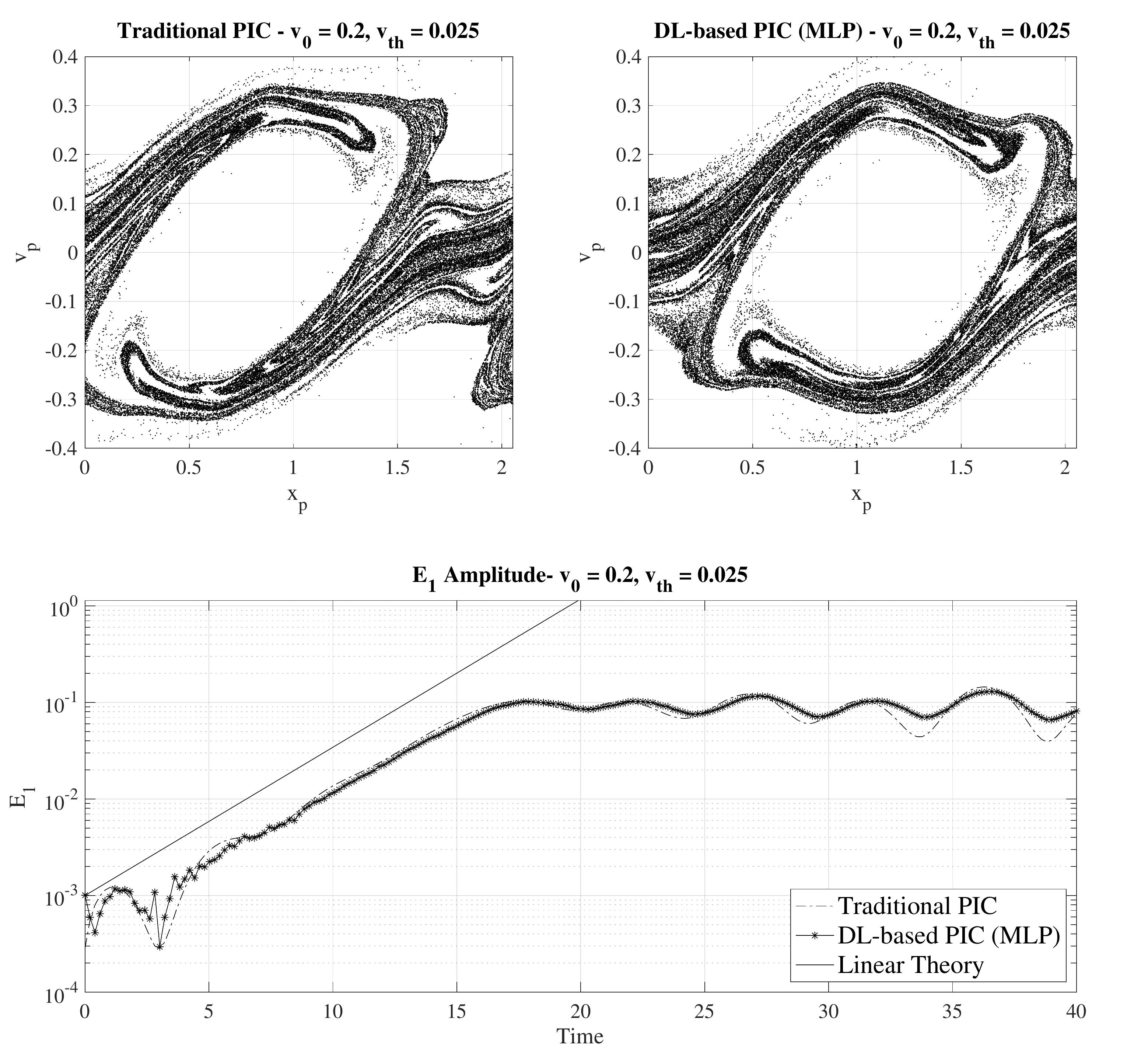}}
\caption{Electric field amplitude of the most unstable mode with traditional PIC, DL PIC and analytical calculations.}
\label{img_E1}
\end{figure}

As part of this work, we then test the DL electric field solver in the proposed PIC method. In addition, we found that the DL-based PIC method using MLP showed better energy and momentum conservation properties. For this reason, we focus on studying the a DL solver using the MLP architecture.

As first step, we validate the results of the DL-based PIC method. We use a configuration that has not been included in the simulations run to generate the training, test and validation data sets. We select a configuration of the electron beams with $v_0 = \pm 0.2$ and $v_{th} = 0.025$.

The top panel of Fig.~\ref{img_E1} shows the electron phase space for traditional PIC method (left panel) and the DL-based PIC method (right panel). Given the fact, the two-stream instability starts at slightly different times, the two phase space plots are not equal but they show similar characteristics, e.g. size of the phase space hole.

To validate the results of the simulation, we compare the DL-based PIC simulation with the results of analytical theory providing us the growth rate of the most unstable mode in the two-stream instability in the cold-beam $v_0 >> v_{th}$ approximation. The bottom panel of Fig.~\ref{img_E1} shows the electric field amplitude of the most unstable mode, $E_1$, during the two-stream instability simulation. The solid line represents the slope predicted by the linear theory (analytical results). The $-.$ and $-*$ lines represent $E_1$ in the traditional and DL-based PIC methods. In the linear phase of the instability (the first phase of the simulation when $E_1$ grows exponentially), the line slopes match the slope predicted by the analytical theory.

\begin{figure}[tb]
\centering
\centerline{\includegraphics[width=0.7\columnwidth]{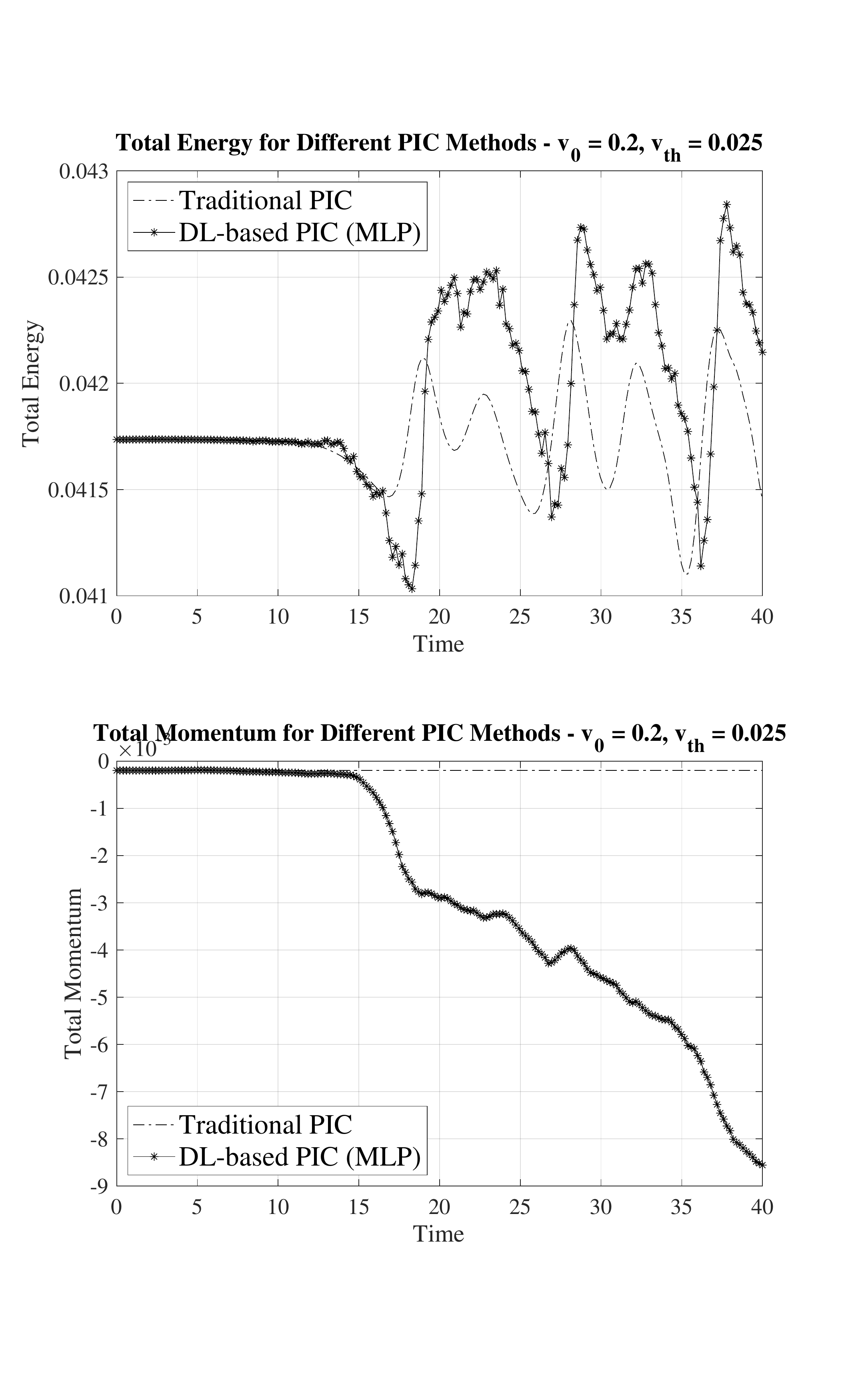}}
\caption{A comparison of total energy and momentum evolution in the traditional and DL-based PIC simulations of the two-stream instability.}
\label{img_energymomenta}
\end{figure}

As mentioned in Section~\ref{background}, it is important to monitor how accurately the total energy and momentum are conserved in the PIC simulations. The top panel of Fig.~\ref{img_energymomenta} shows the variation of the total energy for the traditional and DL-based PIC. In both methods, the total energy is not conserved with maximum variation of approximately 2 \%. The bottom panel shows the total momentum evolution during the two-stream instability. While the traditional PIC simulation conserves the momentum, the momentum decreases in the DL-based PIC simulation. 

\begin{figure}[tb]
\centering
\centerline{\includegraphics[width=\columnwidth]{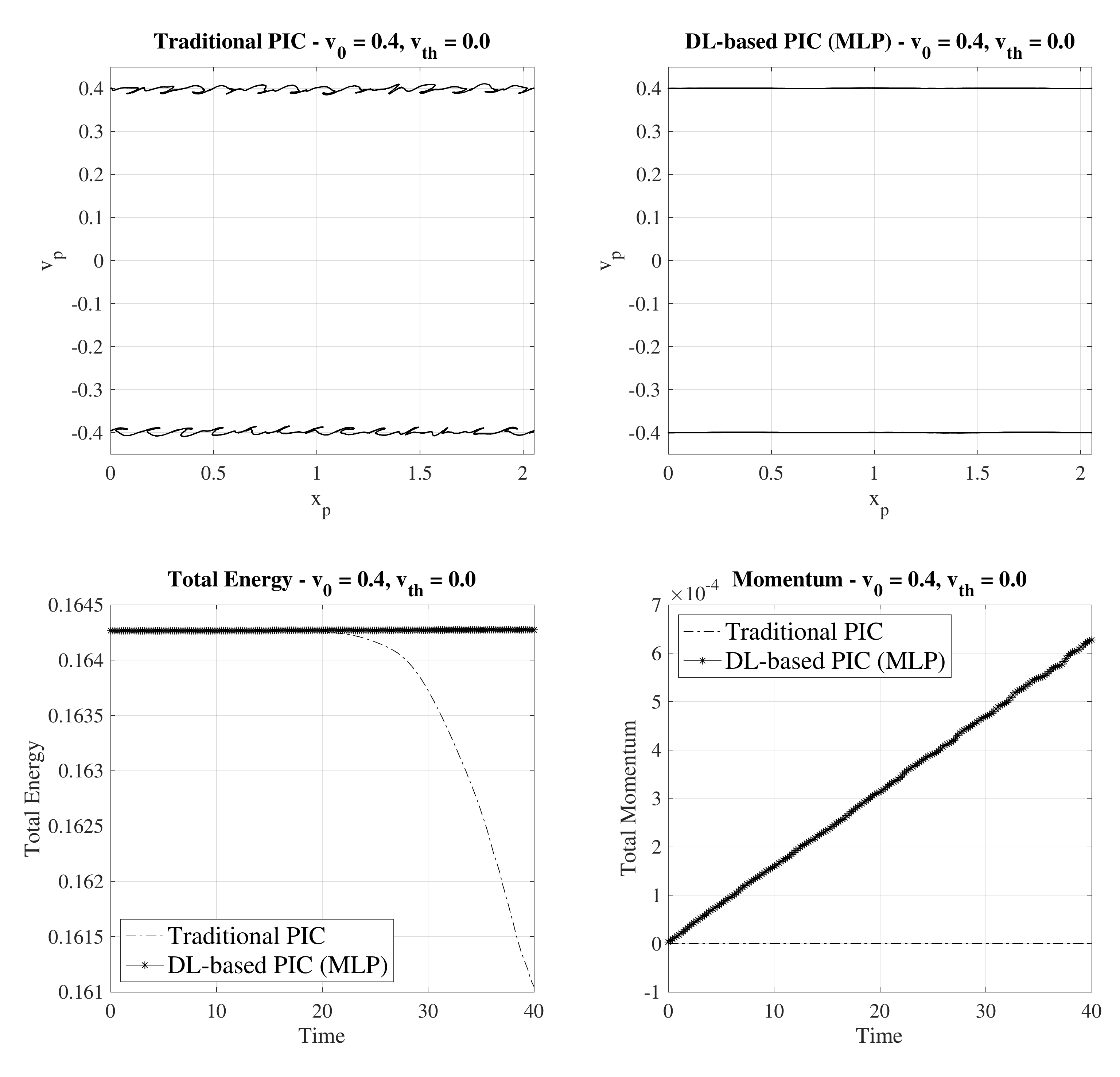}}
\caption{The top panels shows the phase space for the traditional and DL-based PIC simulations for two beams at $v_0 = \pm 0.4$. The traditional PIC simulation is affected by the cold-beam instability. The bottom panels presents the total energy and momentum for the two methods.}
\label{img_v04}
\end{figure}
 
 As the last experiment, we simulate two electron beams with $v_0= \pm 0.4$ and $v_{th}=0.0$. This configuration system is stable against the two-stream instability and the two beams should continue streaming. The phase space at the end of simulation ($t = 40$) in the traditional and DL-based PIC methods is shown in top panels of Fig.~\ref{img_v04}. The phase space obtained with the traditional PIC exhibits a nonphysical behavior, visible in ripples in the phase space. This numerical instability is called \emph{cold-beam instability} and affects momentum and energy conserving PIC methods~\cite{birdsall2018plasma}. The DL-based simulation does not seem affected by the \emph{cold-beam instability}. The numerical instability is visible in the non-conservation of the total energy for the traditional PIC method (bottom left panel Fig.~\ref{img_v04}). While the numerical instability is not present in the DL-based PIC method, the variation of the total momentum increases with the simulation evolution.

\section{Related Work}
\label{related}
The PIC method is a workforce in computational plasma physics. The different PIC algorithm formulations are explained in detail in two classic textbooks~\cite{birdsall2018plasma,hockney2021computer}. Several PIC formulations have been proposed to extend PIC capabilities in terms of robustness against numerical instabilities, energy conservation~\cite{markidis2011energy} and coupling fluid with PIC codes to model large scale systems~\cite{markidis2014fluid,markidis2018polypic}. While Machine Learning has been been applied to study the results of plasma simulations~\cite{markidis2020automatic,innocenti2021unsupervised}, still DL approaches have not found a main role in the PIC methods.

\section{Discussion and Conclusions}
\label{conclusion}
This paper presented a new PIC method using DL to predict the electric field on a grid given the particle phase space information. We trained an MLP and a CNN using the phase space information and associated electric field of the traditional PIC method. We verified that DL-based produces the correct results using the two-stream instability: the DL-based PIC provides the expected growth rate of the two-stream instability.  The DL-based PIC does not conserve the total energy and momentum. However, the DL-based is stable against the cold-beam numerical instability. More studies, such as spectral analysis of errors in the electric field values, are needed to gain more insight into the DL-based PIC methods.

This work is only the first step towards the integration of DL techniques in PIC methods. DL-based PIC methods can be improved in several ways. For instance, more accurate training data sets can be obtained by running Vlasov codes that are not affected by the PIC numerical noise. In this work, we use the NGP interpolation scheme for the phase space bidding. The usage of higher-order interpolation functions would likely improve the performance of the DL electric field solver as it would mitigate numerical artifacts introduced by the binning. Taking into account that phase space and electric field values at a certain time step are very similar to the values in the previous and next time steps, the usage of neural networks fit to encode time sequences, such as Residual networks (ResNet), might be a better fit to DL-based PIC methods than MLPs. 

To be competitive with other PIC methods in terms of physical accuracy, a DL-based PIC should explicitly integrate the conservation laws in the scheme. The DL electric field solver does not use any information from the governing equations and conservation laws, such as total energy and conservation law. Physics-informed Neural Networks (PINN) allow us to encode such equations in the neural network~\cite{raissi2019physics}. The usage of PINN would improve the conservation of total energy and momentum and the performance of the DL-based PIC method.

In this paper, we did not cover the computational performance of the DL-based PIC method. However, we highlight that the DL electric field solver is a simple prediction/inference step involving a series of matrix-vector multiplications. These operations can also be offloaded to GPU systems when using DL frameworks, such as TensorFlow. On the contrary, traditional PIC methods require a linear system that involves more operations than the prediction/inference step. An additional advantage of the DL electric field solver is that it does not need communication when running the DL-based on distributed memory systems as all neural networks can be loaded on each process.

Future work will include a characterization of the DL-based PIC computational performance. We also intend to extend the method to study two- and three-dimensional systems for electromagnetic problems.

\section*{Acknowledgments}
All computation needed for this paper was enabled by resources provided by the Swedish National Infrastructure for
Computing (SNIC), partially funded by the Swedish Research Council through grant agreement no. 2018-05973. S.M. acknowledges funding from the European Commission H2020 program, Grant Agreement No. 801039 (EPiGRAM-HS).

\bibliographystyle{IEEEtran}
\bibliography{main}
\end{document}